\documentclass[usenatbib]{mn2e}
\usepackage{epsfig}
\usepackage{subfigure}
\usepackage{amssymb}
\usepackage{xcolor}

\newcommand{\bn}{\begin{enumerate}}
\newcommand{\en}{\end{enumerate}}
\newcommand{\bi}{\begin{itemize}}
\newcommand{\ei}{\end{itemize}}

\newcommand{\rp}[1]{\left(#1\right)}

\newcommand{\Sigsfr}{\Sigma_{\rm SFR}}
\newcommand{\HI}{H\,{\sc i}}
\newcommand{\NHI}{{N_{\rm HI}}}
\newcommand{\Msun}{M_{\odot}}
\newcommand{\Mstar}{M_{\star}}

\newcommand{\rhosdot}{\dot{\rho}_{\star}}
\newcommand{\rhostar}{\rho_{\star}}

\newcommand{\apj}{ApJ}

\newcommand{\apjl}{ApJL}
\newcommand{\apjs}{ApJS}
\newcommand{\mnras}{MNRAS}
\newcommand{\aj}{AJ}
\newcommand{\araa}{ARA\&A}

\newcommand{\pasp}{PASP}

\title[Cosmic Star Formation]{Effects of cosmological parameters and star formation models on the cosmic star formation history in $\Lambda$CDM cosmological simulations}

\author[Choi \& Nagamine]
{Jun-Hwan Choi\thanks{Email: jhchoi@physics.unlv.edu} Kentaro Nagamine\thanks{Visiting Researcher, Institute for the Physics and Mathematics of the Universe, University of Tokyo, 5-1-5 Kashiwanoha, Kashiwa, Chiba 277-8568 Japan} \vspace{0.2cm}\\
Department of Physics and Astronomy, University of Nevada, Las Vegas, 4505 S. Maryland Pkwy, Las Vegas, NV, 89154-4002, U.S.A.}

\begin{document}
\maketitle

\begin{abstract}
We investigate the effects of the change of cosmological parameters and star formation (SF) models on the cosmic SF history using cosmological smoothed particle hydrodynamics (SPH) simulations based on the cold dark matter (CDM) model.
We vary the cosmological parameters within 1-$\sigma$ error from the WMAP best-fit parameters, and find that such changes in cosmological parameters mostly affect the amplitude of the cosmic SF history. 
At high redshift (hereafter high-$z$), the star formation rate (SFR) is sensitive to the cosmological parameters that control the small-scale power of the primordial power spectrum, while the cosmic matter content becomes important at lower redshifts. 
We also test two new SF models: 1) the `Pressure' model based on the work by Schaye \& Dalla Vecchia (2008), and 2) the `Blitz' model that takes the effect of molecular hydrogen formation into account, based on the work by Blitz \& Rosolowsky  (2006). 
Compared to the previous conventional SF model, the Pressure model reduces the SFR in low-density regions and shows better agreement with the observations of the Kennicutt-Schmidt law. 
This model also suppresses the early star formation and shifts the peak of the cosmic SF history toward lower redshift, more consistently with the recent observational estimates of cosmic SFR density. 
The simulations with the new SF model also predict lower global stellar mass densities at high-$z$, larger populations of low-mass galaxies and a higher gas fraction in high-$z$ galaxies.
Our results suggest that there is room left in the model uncertainties to reconcile the discrepancy that was found between the theory and observations of cosmic SF history and stellar mass density.
Nevertheless, our simulations still predict higher stellar mass densities than most of the observational estimates. 
\end{abstract}


%% ==================================================================

\section{Introduction}
\label{sec:intro}

Star formation is a fundamental and important physical process for understanding galaxy formation.
Stars emit a large fraction of the observed photons, and their formation and evolution significantly influence galaxy evolution.
Over the past decade, various high-$z$ observations started to unveil the rough history of cosmic star formation \citep[e.g.,][and references therein]{Madau.etal:96, Lilly.etal:96, Steidel.etal:96, Dickinson.etal:03, Giavalisco.etal:04, Ouchi.etal:04, Hopkins.Beacom:06, Bouwens.etal:07, Yan.etal:08, Bouwens.etal:09}.
There are yet considerable uncertainties in the dust extinction correction, the faint-end slope of the luminosity function, and the stellar initial mass function (IMF).
However, these observations show a rough picture of the cosmic SFR density ($\rhosdot$), which gradually increases from high-$z$ to $z \sim 4$, peaks at $z \approx 2-4$, and then rapidly declines from $z\sim 2$ to $z=0$. 
It is then of significant interest to see if the standard model of cosmic structure formation based on the CDM model can reproduce the observed features of cosmic SF history, and to further develop insight into the physical processes which govern the cosmic star formation.

Numerical simulations have been used to investigate the cosmic star formation history \citep[e.g.,][]{Cen.Ostriker:92,Katz.etalL:96,Yepes.etal:97,Nagamine.etal:00, Weinberg.etal:02,Springel.Hernquist:03_SFR,Nagamine.etal:04,Nagamine.etal:06,Dave:08}, however, most theoretical predictions remain somewhat uncertain due to the following reasons.  
First of all, the physics of star formation and its feedback have not been clearly understood yet, because they are inherently complex problems, involving nonlinear dynamics, radiative and chemical processes on wide range of scales \citep[e.g.,][]{McKee.Ostriker:07}.
Secondly, the resolution limitation of simulations forbids the detailed modeling of SF and its feedback from first principles, and cosmological simulations must rely on empirical SF laws \citep[e.g,][]{Schmidt:59, Kennicutt:98a,Kennicutt:98b}.
The simulation results could depend on the details of the adopted models of SF and its feedback, and there are still significant freedoms in the formulation of these models. 

Originally, numerical cosmologists studied the cosmic SF history with the motivation to use it as a probe of cosmological models, and examine the effects of largely different cosmological models on $\rhosdot$ \citep[e.g.,][]{Nagamine.etal:00, Weinberg.etal:02}. 
This is because different cosmological models result in different power spectra of matter density fluctuations, which would be reflected as different shapes of $\rhosdot(z)$.  

Since then, the situation has dramatically changed, and we now know the values of our cosmological parameters within $\sim$10\% accuracy \citep{Komatsu.etal:09, Komatsu.etal:10}, and the resolution of numerical simulations has also improved significantly since late 1990's.
Back then, cosmological hydro simulations with 64$^3$ particles were standard, but now we can go up to $300^3 - 400^3$ particles.
Given these advancements, it would be worthwhile to quantify the changes in $\rhosdot(z)$ in more detail when we deviate the cosmological parameters by 1-$\sigma$ error from the WMAP best-fit values \citep{Komatsu.etal:09, Komatsu.etal:10}.
An interesting recent change is that the values of $\sigma_8$ and the power-index of primordial power spectrum $n_s$ have become smaller with $\sigma_8 < 0.9$ and $n_s < 1.0$.  
The goal of this paper is to quantify the dependence of $\rhosdot(z)$ on cosmological parameters and the adopted SF model in detail, as it has not been documented in the literature yet. 

A very comprehensive work was performed by \citet[][hereafter SH]{Springel.Hernquist:03_SFR}, in which they estimated $\rhosdot(z)$ using a large set of cosmological SPH simulations with a novel subgrid model for multiphase interstellar medium (ISM) \citep[][]{Springel.Hernquist:03}.
The parameters in their SF model were constrained by the observed Kennicutt-Schmidt law \citep{Kennicutt:98a,Kennicutt:98b}.  
They clearly showed how $\rhosdot(z)$ depends on the numerical resolution, but they did not examine the effects of different cosmologies and SF models in detail. 

Based on the work of SH, \citet[][hereafter HS model]{Hernquist.Springel:03} provided a simple, analytic reasoning to identify physical processes that drive the evolution of $\rhosdot(z)$. 
They argued that the early phase of cosmic star formation is driven by the gravitational structure formation, and provided an analytic formula for $\rhosdot(z)$, which includes the Hubble parameter $H(z)$ and the SFR function as a function of halo mass.  
Compared to the observational estimates, the HS model with the WMAP cosmology predicts a higher $\rhosdot$ at high-$z$ with a peak at $z\approx 5-6$, resulting in higher stellar mass densities ($\Omega_{\ast}$) at $z\ge 1$ than the observations. 
Later, \citet{Nagamine.etal:04} compared two different kinds of numerical simulations (Eulerian TVD and SPH) with observations, and found that the numerical simulations generally predicted higher SFR at high-$z$.  

Although the SH model provides a novel description of star formation and its feedback, this model is still not perfect. 
As mentioned above, the star formation in the SH model is constrained by the Kennicutt-Schmidt law which is the correlation between the average star formation surface density ($\rm \Sigma_{SFR}$) and average total gas surface density ($\rm \Sigma_{gas}$).
Recent observations, which spatially resolve star formation in a given galaxy, show that $\Sigsfr(r)$ is a function of the $\rm \Sigma_{{\rm H_2}}(r)$, and  $\rm \Sigma_{gas}(r)/\Sigma_{{\rm H_2}}(r)$ is not constant.  
This suggests that $\Sigsfr$ is indeed a function of molecular hydrogen surface density ($\rm \Sigma_{{\rm H_2}}$), rather than $\rm \Sigma_{gas}$ \citep{Wong.Blitz:02,Heyer.etal:04,Blitz.Rosolowsky:06,Bigiel.etal:08}.
Therefore, it will be important to take into account the effect of ${\rm H_2}$ density in our SF model.  
In this paper, we examine new SF models which consider the contribution of ${\rm H_2}$, and test the effects on $\rhosdot(z)$ and star formation on small scales. 

This paper is organized as follows. In Section \ref{sec:method}, we describe our simulations with a focus on the new SF models.
In Section \ref{sec:comp}, we present the effects of differing cosmological parameters on the cosmic SF history.
In Section \ref{sec:psf}, we compare the two new SF models with the SH model. 
The effect of new SF models on $\rhosdot(z)$ and galaxy formation is investigated in Section \ref{sec:new}.
Finally we summarize our findings in Section \ref{sec:summary}.

\section{Numerical technique}
\label{sec:method}

We use the updated version of the Tree-particle-mesh (TreePM) smoothed particle hydrodynamics (SPH) code GADGET-2 \citep{Springel:05}.
Our conventional code includes radiative cooling by H, He, and metals \citep{Choi.Nagamine:09}, heating by a uniform UV background of a modified \citet{Haardt.Madau:96} spectrum \citep{Katz.etal:96,Dave.etal:99}, star formation, supernova feedback, a phenomenological model for galactic winds, and a sub-resolution model of multiphase ISM \citep{Springel.Hernquist:03}. 
In this multiphase ISM model, the high-density ISM is pictured to be a two-phase fluid consisting of cold clouds in pressure equilibrium with a hot ambient phase.
Cold clouds grow by radiative cooling out of the hot medium, and this material forms the reservoir of baryons available for star formation. 
Since the details of the treatment are described by both SH and \citet{Choi.Nagamine:09}, we do not repeat them here.

\subsection{Cosmological parameters}
\label{sec:ic_cosmo}

In this paper, we adopt the following fiducial cosmology which is consistent with the latest WMAP result: $\Omega_m = 0.26$, $\Omega_{\Lambda} = 0.74$,  $\Omega_b = 0.044$, $h=0.72$, $n_{s}=0.96$, and $\sigma_{8}=0.80$.
In order to see the effect of each cosmological parameter on $\rhosdot(z)$, we vary one parameter at a time from the fiducial model while the other parameters are kept fixed.
Except $\sigma_{8}$, we vary each parameter from the fiducial value by approximately one sigma error in the WMAP result \citep{Komatsu.etal:09, Komatsu.etal:10}.
The variation of $\sigma_{8}$ is chosen to be 0.1 to include the previously popular $\sigma_{8}$ = 0.9 in our parameter set.
The tested cosmological models are in the range of $n_s = 0.92 - 1.0$, $\sigma_{8} = 0.7 - 0.9$, $\Omega_{b} = 0.038 - 0.05$,  $(\Omega_{m}, \Omega_{\Lambda}) = (0.24, 0.76) \; {\rm and} \; (0.28,0.72)$ \footnote{The flat universe is one of the most robust constraint of WMAP, therefore we fix $\Omega_{m} + \Omega_{\Lambda} = 1$ for this variation.}. 
The runs used in this comparison are listed in Table~\ref{table:sim1}.

\begin{table*}
\begin{center}
\begin{tabular}{cccccccc}
\hline
Runs & $\Omega_{m}$ & $\Omega_{\Lambda}$ & $\Omega_{b}$ & $n_s$ & $\sigma_8$ & $m_{\rm DM}$ & $m_{\rm gas}$ \\ 
\hline
\hline
Fiducial              & 0.26 & 0.74 & 0.044 & 0.96 & 0.8 & $5.96 \times 10^6$ & $1.21 \times 10^6$ \cr 
High $n_s$            & 0.26 & 0.74 & 0.044 & 1.00 & 0.8 & $5.96 \times 10^6$ & $1.21 \times 10^6$ \cr
Low  $n_s$            & 0.26 & 0.74 & 0.044 & 0.92 & 0.8 & $5.96 \times 10^6$ & $1.21 \times 10^6$ \cr
High $\sigma_8$       & 0.26 & 0.74 & 0.044 & 0.96 & 0.9 & $5.96 \times 10^6$ & $1.21 \times 10^6$ \cr
Low  $\sigma_8$       & 0.26 & 0.74 & 0.044 & 0.96 & 0.7 & $5.96 \times 10^6$ & $1.21 \times 10^6$ \cr
High $\Omega_{m}$     & 0.24 & 0.76 & 0.044 & 0.96 & 0.8 & $5.41 \times 10^6$ & $1.21 \times 10^6$ \cr
Low  $\Omega_{m}$     & 0.28 & 0.72 & 0.044 & 0.96 & 0.8 & $6.51 \times 10^6$ & $1.21 \times 10^6$ \cr
High $\Omega_{b}$     & 0.26 & 0.74 & 0.038 & 0.96 & 0.8 & $6.12 \times 10^6$ & $1.05 \times 10^6$ \cr
Low  $\Omega_{b}$     & 0.26 & 0.74 & 0.050 & 0.96 & 0.8 & $5.79 \times 10^6$ & $1.38 \times 10^6$ \cr
\hline
\end{tabular}
\caption{
The simulations employed in the cosmological parameter test.
All the simulations are N216L10 series whose general properties are shown in Table~\ref{table:sim2}.
The mass resolution varies slightly due to the variation of the cosmic matter content
}
\label{table:sim1}
\end{center}
\end{table*}

\begin{table*}
\begin{center}
\begin{tabular}{ccccccc}
\hline
Series &  Box-size & ${N_{\rm p}}$ & $m_{\rm DM}$ & $m_{\rm gas}$ & $\epsilon$ & $z_{\rm end}$ \\
\hline
\hline
N216L10    & 10.0  & $2\times 216^3$  & $5.96 \times 10^6$ & $1.21 \times 10^6$ &  1.85  & 2.75 \cr
\hline
N400L34   & 33.75 & $2\times 400^3$  & $3.49 \times 10^7$ & $7.31 \times 10^6$ &  3.375  & 1.0 \cr
\hline
N400L100   & 100.0 & $2\times 400^3$  & $9.12 \times 10^8$ & $1.91 \times 10^8$ &  6.45  & 0.0 \cr
\hline
\end{tabular}
\caption{
The two series of simulations employed for the comparison study of star formation models.
The box-size is given in units of $h^{-1}$Mpc, ${N_{\rm p}}$ is the particle number of dark matter and initial gas (hence $\times\, 2$), $m_{\rm DM}$ and $m_{\rm gas}$ are the masses of dark matter and gas particles in units of $h^{-1}M_{\sun}$, respectively, $\epsilon$ is the comoving gravitational softening length in units of $h^{-1}$kpc, and $Z_{\rm end}$ is the ending redshift of the simulation.
The value of $\epsilon$ is a measure of spatial resolution.
Note that the star particle mass is a half of the initial gas particle mass.
We implement three star formation models (the SH model, the Blitz model, and the Pressure model) for the N216L10 series and two star formation models (the SH model and the Pressure model) for the N400L100 series.
}
\label{table:sim2}
\end{center}
\end{table*}

Star formation at high redshift takes place mostly in low-mass galaxies, and it progressively shifts to massive systems at lower redshifts.
We need a high-resolution simulation to resolve the star formation in low-mass galaxies and a large box size to include a large number of massive galaxies, which requires substantial computational resources.
In order to get around this difficulty, we employ a large number of runs with different resolution and volumes. 
Because of this, we limit the particle number of our fiducial runs to $2\times 216^{3}$ (gas + dark matter) particles in a comoving box of $(10\,h^{-1}{\rm Mpc})^3$ (hereafter the N216L10 series). 
We stop this series at $z=2.75$, as it misses the long wavelength perturbations at lower redshifts. 
The resolution of N216L10 series is not adequate to properly simulate the entire cosmic SF history, but it is sufficient to show the differences due to the variation of the cosmological parameters.

\subsection{Star formation models}
\label{sec:sfmodel}

\noindent
$\bullet$ {\bf SH model:}\\
In the multiphase ISM model of SH, the star formation is modelled as follows.
If the dense gas is Jeans unstable and is rapidly cooling, a fraction of the gas mass is converted into a star particle with a rate 
\begin{eqnarray}
{\dot{\rho_{\star}}} = (1-\beta)\rho_c / t_{\rm SFR}, 
\label{equ:SH}
\end{eqnarray}
where $\rho_c$ is the gas density of the cold cloud, and $\beta$ is the mass fraction of high-mass stars that instantly die as supernovae, determined by the stellar initial mass function.
The star formation time-scale $t_{\rm SFR}$ is taken to be proportional to the local dynamical time of the gas: $t_{\rm SFR}(\rho) = t_0^{\star}\rp{\rho / \rho_{\rm th}}^{-1/2},$ where the value of $t_0^{\star}=2.1\,{\rm Gyr}$ is chosen in isolated disk galaxy simulations to match the Kennicutt-Schmidt law: 
\begin{eqnarray}
\Sigsfr = \left\{ \begin{array}{ll} 0 & \mbox {if $\Sigma_{\rm gas} < \Sigma_{\rm th}$} \\
   A(\Sigma_{\rm gas}/1 \Msun\,{\rm pc}^{-2})^n & \mbox {if $\Sigma_{\rm gas} > \Sigma_{\rm th}$,} \end{array} \right. 
\label{equ:KS}
\end{eqnarray}
where $\Sigma_{\rm th}$ is the SF threshold surface density.
Observations suggest that $A = 2.5 \pm 0.7\,\Msun\,{\rm yr}^{-1}\,{\rm kpc}^{-2}$, $n=1.4 \pm 0.15$, and $\Sigma_{\rm th} \sim 10\,\Msun$\,pc$^{-2}$ \citep{Kennicutt:98a,Kennicutt:98b}.

The supernova explosions add thermal energy to the hot phase of the ISM and evaporate cold clouds. 
This is described as ${\dot{\rho_c}} = C\beta \rho_c / t_{\star}$, where the feedback efficiency parameter ``$C$'' has the density dependence $C(\rho) = C_0\rp{\rho / \rho_{\rm th}}^{-4/5},$ following \citet{McKee.Ostriker:77}.
This evaporation process of cold clouds establishes a tight self-regulation mechanism for star formation in the ISM, where the ambient hot medium quickly evolves toward an equilibrium temperature.

Although the SH model has shown significant improvements over the earlier SF models, it is still incomplete.
We know from observations that star formation takes place in the molecular clouds.  Therefore the next natural step is to implement alternative SF models which include the effect of molecular hydrogen (${\rm H_2}$), and study the consequences on the cosmic star formation history \citep[e.g.,][]{Gnedin.etal:09}. 
Unfortunately, current cosmological simulations cannot resolve the detailed structure of molecular clouds while solving the formation of thousands of galaxies on a  $\gtrsim$10\,Mpc scale, so we need a model to estimate the ${\rm H_2}$ fraction from the total gas density. 
Here, we consider two new SF models and compare them with the original SH model.

\vspace{0.3cm}
\noindent
$\bullet$ {\bf Blitz model:}\\
This model is based on the ${\rm H_2}$ density-pressure relation derived by \citet{Blitz.Rosolowsky:06}.
They argued that the mean ratio of molecular to atomic hydrogen surface density is related to the interstellar gas pressure as follows: $\Sigma_{\rm H_2} / \Sigma_{\rm HI} = (P_{\rm ext}/P_0)^{\alpha}$, where $P_{\rm ext}$ is the interstellar gas pressure, $P_{0}/k = 4.3 \times 10^{4}$\,cm$^{-3}$\,K, and $\alpha \approx 0.92$.
Using this relationship, we can compute the amount of ${\rm H_2}$ from the total gas density and pressure. 
The projected SFR density in this model is
\begin{eqnarray}
\Sigsfr = 0.1 \epsilon \Sigma_{\rm gas} \left [1 + \left(\frac{P_{\rm ext}}{P_{0}}\right)^{-\alpha} \right]^{-1},
\label{equ:BSF}
\end{eqnarray}
where $\epsilon \approx 10 - 13 \, {\rm Gyr}^{-1}$.
\citet{Blitz.Rosolowsky:06} argued that this relationship recovers the observed SFR surface density better than the \citet{Kennicutt:98a} law, especially in the low density regime, where the molecular density is lower than the H{\sc i} density and $P_{\rm ext} < P_{0}$.

We assume that we can replace the surface density with the 3-dimensional density, and rewrite the above equation as
\begin{eqnarray}
\rhosdot = \rho_{\rm gas} \left [1 + \left(\frac{P_{\rm ext}}{P_{0}}\right)^{-\alpha} \right]^{-1} {\rm Gyr}^{-1},
\label{equ:BSF2}
\end{eqnarray}
after adopting $\epsilon =10\,{\rm Gyr}^{-1}$ as suggested in \citet{Blitz.Rosolowsky:06}.
We note that this SF law is very similar to the one adopted by \citet{Kravtsov:03} except for the pressure term in the bracket. 
\citet{Kravtsov:03} assumed $\rhosdot = \rho_{\rm gas} / \tau_\star$ with $\tau_\star = 4$\,Gyr. 

When $P_{\rm ext} > P_{0}$, the molecular surface density becomes greater than the atomic hydrogen surface density. 
In this high-pressure regime, we assume that the SF law reverts to the Kennicutt-Schmidt law, and adopt equation~(\ref{equ:KS}).
We apply equation~(\ref{equ:BSF2}) only in the low-pressure regime with $P_{\rm ext} < P_{0}$.
We call this new model the `Blitz' model.

\vspace{0.3cm}
\noindent
$\bullet$ {\bf Pressure model:}\\
This model explicitly formulates the conversion between gas surface density ($\Sigma_{\rm gas}$) and gas volume density ($\rho_{\rm gas}$).
Previously, we assumed $\Sigma_{\rm gas}/\Sigsfr  = \rho_{\rm gas}/\rhosdot$, which is only true if the disk scale-height is constant or the equation of state (EOS) behaves as $P \propto \rho^2$.
\citet{Schaye:01} and \citet{Schaye.DallaVecchia:08} proposed the ``Jeans column density'', and argued that the scale-height will be of the order of local Jeans scale for self-gravitating discs, because the density typically fluctuates on the local Jeans scale.

One may argue that this model takes into account the effect of H$_2$ cooling better than the previous SF models based on the three-dimensional gas density, because the disk instability leads to the collapse of molecular clouds and star formation occurs within them due to the H$_2$ cooling. 
Owing to the current limitation in computational power, it is still impossible to simulate the formation of molecular clouds explicitly in cosmological simulations.
Therefore considering the Jeans instability in a disk could be a useful approximation to take account of the effect of disk instability, which leads to the molecular cloud formation, in cosmological simulations. 

Using the Jeans column density $\Sigma_{g,J}$, surface and volume gas densities are related as follows:
\begin{eqnarray}
\Sigma_{\rm gas} \sim \Sigma_{g,J} & \equiv & \rho_{\rm gas} L_{J}  =  \sqrt{\frac{\gamma}{G} f_{g} P_{tot}} ,
\label{equ:Jean}
\end{eqnarray}
where $L_J = c_s / \sqrt{G \rho_{tot}}$ is the Jeans length, $c_s = \sqrt{\gamma P_{tot}/\rho_{\rm gas}}$ is the local sound speed, $f_g$ is the  mass fraction in gas (i.e., $\rho_{\rm gas} = f_g \rho_{tot}$), and $f_{th}$ is the fraction of mid-plane pressure that is thermal (i.e., $P = f_{th} P_{tot}$, where $P$ is the thermal pressure).

Throughout our calculation, we assume $f_g = f_{th}$, therefore $f_{g}P_{tot} = (f_{g}/f_{th}) P = P$.
From equations~(\ref{equ:KS}) and (\ref{equ:Jean}), we can derive the new SF time-scale:
\begin{eqnarray}
t_{\rm SFR} =  \frac{\Sigma_{\rm gas}}{\Sigsfr} =  A^{-1} 
\left(1 \Msun\,{\rm pc}^{-2} \right)^n \left( 
\frac{\gamma}{G} P \right)^{(1-n)/2} ,
\label{equ:tsfr}
\end{eqnarray}
where we adopt $n = 1.4$ and $\gamma = 5/3$ as the default values.
In this SF model, we keep the multiphase ISM model of SH. 

We compute the SF threshold density as $\rho_{th} = G/f_g (\Sigma_{th}/c_s)^{2}$. 
To compute $c_s$ for the gas at the threshold density, we need to know the temperature and the mean atomic weight of the gas.
Here, we assume $T_{th} = 500 K$ and $\mu_{th} = 1.2$, because star formation occurs in the cold gas.
The resulting number density is $n_{th} \sim 1.2 h^{2} {\rm cm}^{-3}$, which is a factor of three higher than the original value used in the SH model.  
As we will discuss later, our new value of $n_{th}$ reproduces the proper threshold column density of star formation, as observed by \citet{Kennicutt:98a}.

\begin{figure*}
\centerline{\includegraphics[width=0.9\textwidth,angle=0] {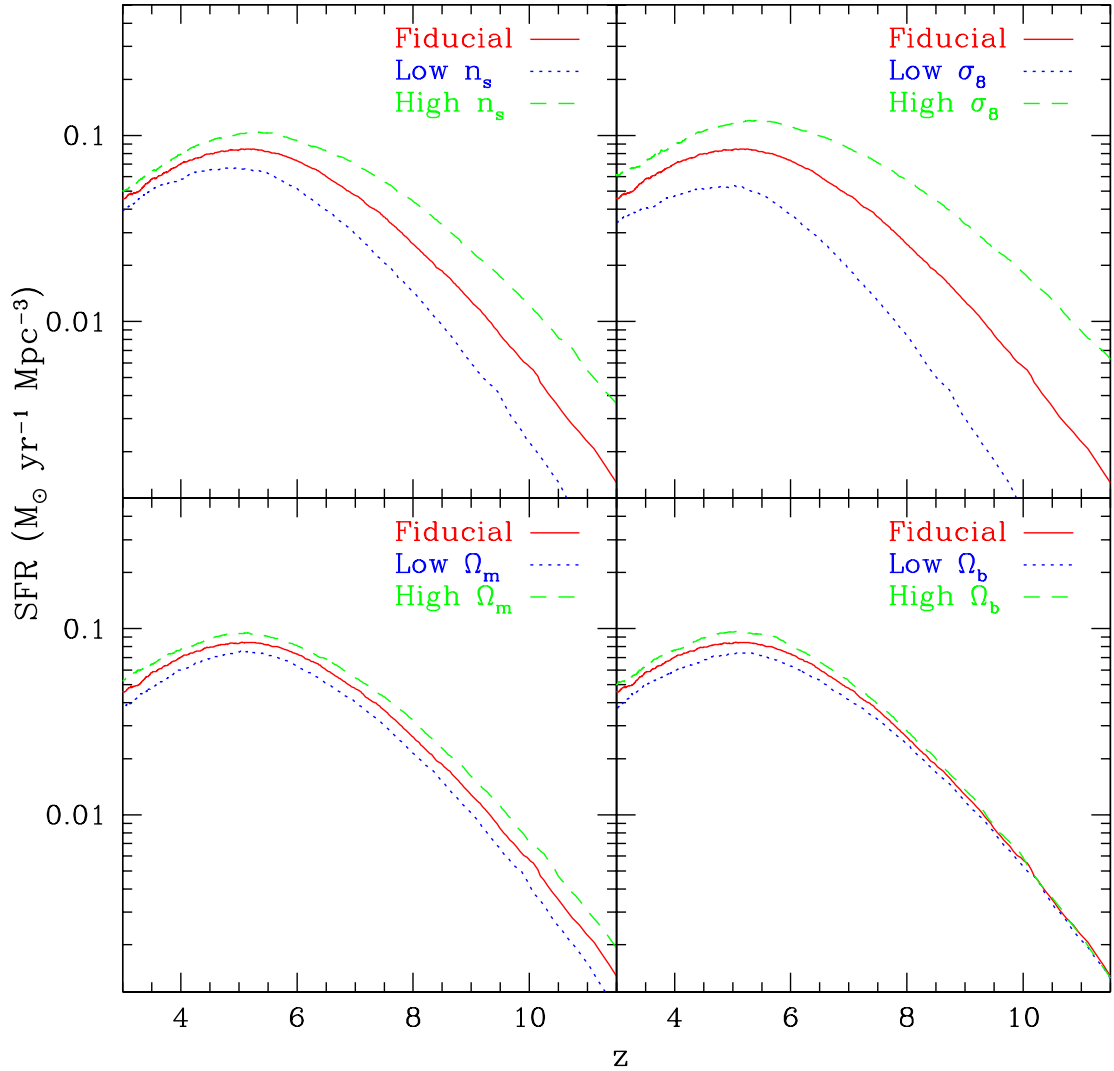}}
\caption{
The cosmic star formation history from the N216L10 series of simulations with different cosmological parameters (see Table~\ref{table:sim1}).
The shape of the cosmic star formation history is not changed by the different cosmological parameters, but the amplitude of the star formation history is significantly affected.
}
\label{fig:comp_w5}
\end{figure*}

The shape of the EOS is closely related to the feedback model. 
In the multiphase prescription of SH, the thermal energy from SN feedback pressurizes the ISM and makes the EOS steeper. \citet{Robertson.etal:04} later derived a fitting formula for this steep, effective EOS of the SH model. 
Although it is possible to use the effective EOS in the simulation for the star-forming regions, we prefer to adopt a polytropic EOS for the star-forming gas to prevent an artificial fragmentation discussed below. 
Note that we still apply the original EOS from the SH model for the non-star forming gas.
\citet{Schaye.DallaVecchia:08} discussed the use of a polytropic EOS,
\begin{eqnarray}
P = K \rho_{\rm gas}^{\gamma_{eff}}.
\label{equ:eos}
\end{eqnarray}
Note that the effective polytropic index $\gamma_{eff}$ is not the same as the usual adiabatic index $\gamma$.
If $\gamma_{eff} = 4/3$, the Jeans mass is independent of the gas density.
If the Jeans mass decreases with density, it will lead to the artificial fragmentation of gas \citep{Bate.Burkert:97}. 
Therefore a polytropic index $\gamma_{eff} = 4/3$ prevents the artificial fragmentation while allowing the collapse to proceed, and we adopt this value as our default. 
The value of $K$ in equation~(\ref{equ:eos}) is computed by inserting this equation into equation~(\ref{equ:Jean}), $K = (G/\gamma) \rho_{th}^{-\gamma_{eff}} \Sigma_{th}^{2}$.

With the new SF time-scale, new density threshold, and new EOS for the star-forming gas, we go back to equation~(\ref{equ:SH}) and complete the new SF model.  
We call this model the `Pressure' model.
In Section \ref{sec:psf}, we will compare $\rhosdot(z)$ from these three SF models using a series of simulation sets presented in Table~\ref{table:sim2}.

\section{Varying the cosmological parameters}
\label{sec:comp}

\begin{figure*}
\centerline{\includegraphics[width=0.9\textwidth,angle=0] {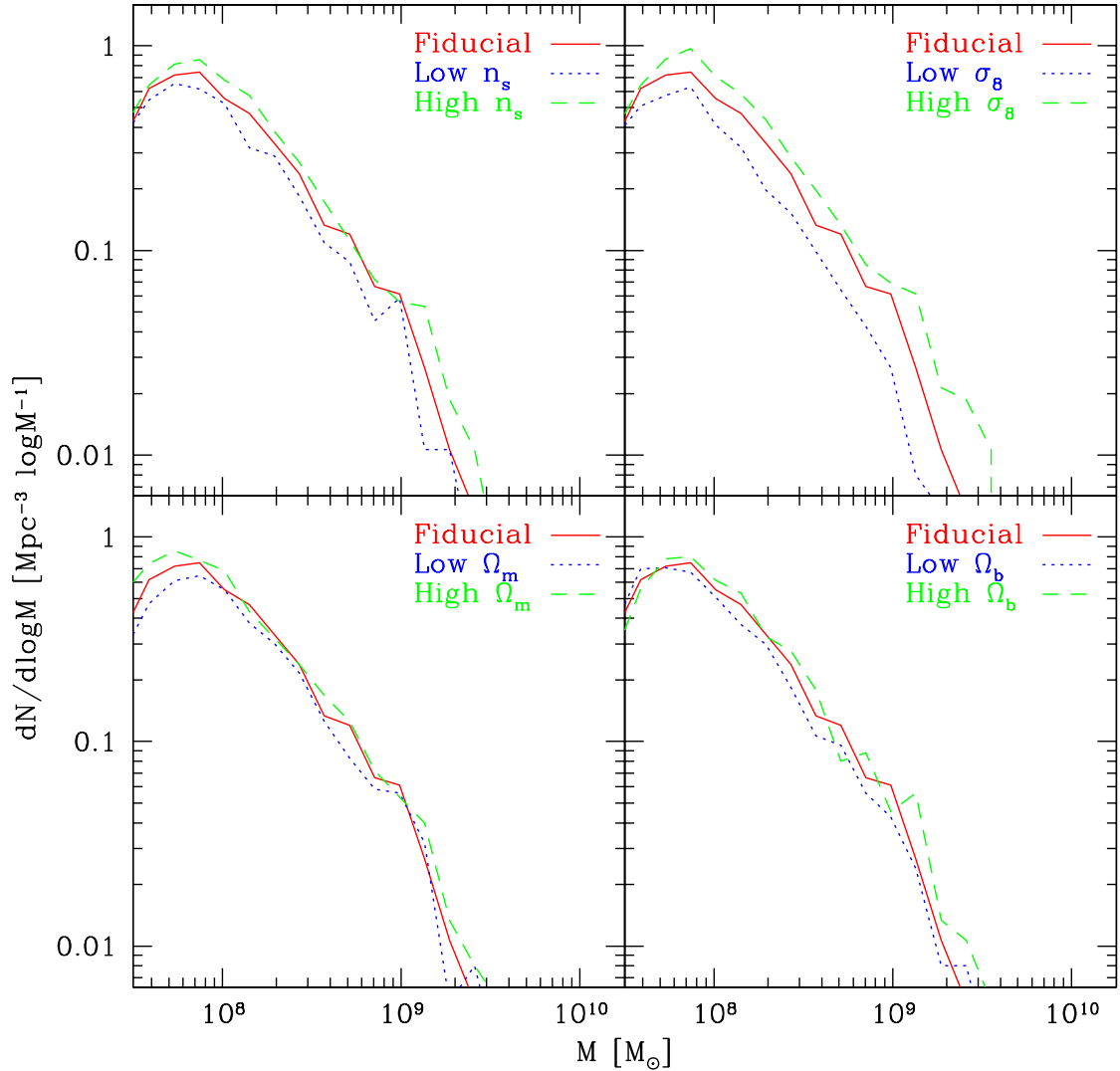}}
\caption{
Galaxy stellar mass functions at $z=3$ for the runs with different cosmological parameters (see Table~\ref{table:sim1}).
The shape of the mass functions is hardly affected by the variation of cosmological parameters, while the amplitude of them is slightly changed by the variations.
}
\label{fig:MF_w5}
\end{figure*}

Figure~\ref{fig:comp_w5} shows the cosmic star formation histories of the N216L10 series with different cosmological parameters.
The general shape of $\rhosdot(z)$ is preserved when the cosmological parameters are varied, but the amplitude changes. 
Figure~\ref{fig:comp_w5} shows that the effect of the primordial power spectrum on $\rhosdot$ is more significant than that of the global mass content of the Universe.
The variance of $\rhosdot$ is quite large at high redshift when $n_s$ and $\sigma_{8}$ are varied. 
This is because the values of $n_s$ and $\sigma_{8}$ determine the primordial power spectrum, which governs the early structure formation.
The global matter content ($\Omega$) also changes the power spectrum, but its contribution is mostly on the transfer function and the growth factor, whose influence becomes more important at lower redshifts.
Therefore, it is expected that the effect of change in $n_s$ and $\sigma_{8}$ is more significant than that of $\Omega$ at high-$z$.

Figure~\ref{fig:comp_w5} also shows that the variance of $\rhosdot$ becomes smaller at lower redshifts when $n_s$ and $\sigma_{8}$ are varied. 
At early times, the cosmic density is high, and the cooling time is short, therefore most dark matter halos can cool the gas and form stars.
During this epoch, the cosmic SFR is mostly driven by the gravitational growth of dark matter halos.
At lower redshifts, the cooling time becomes longer due to decreasing cosmic density and hotter intergalactic medium (IGM).   
Then the effect of gas physics starts to play a more important role in determining the SFR, and the tight coupling between dark matter halo growth and SFR becomes weaker.

The cosmological matter densities ($\Omega_{\Lambda}$, $\Omega_{m}$, and $\Omega_{b}$) contribute to the cosmic SF history in two different ways.
The values of $\Omega_{\Lambda}$ and $\Omega_{m}$ control the redshift evolution of the Hubble parameter, $H(z)$, which  affects the linear growth factor of structure formation and the expansion rate of the universe. 
Therefore the changes in $\Omega_{\Lambda}$ and $\Omega_{m}$ affect the $\rhosdot(z)$ through the growth factor.
At low redshift, the decrease of the mean density of the Universe results in the declining efficiency of gas cooling \citep{White.Frenk:91}.
Hence, $\Omega_{\Lambda}$ and $\Omega_{m}$ continuously influence the cosmic star formation history, which is confirmed in Figure~\ref{fig:comp_w5}.

\begin{figure*}
\centerline{\includegraphics[width=0.9\textwidth,angle=0] {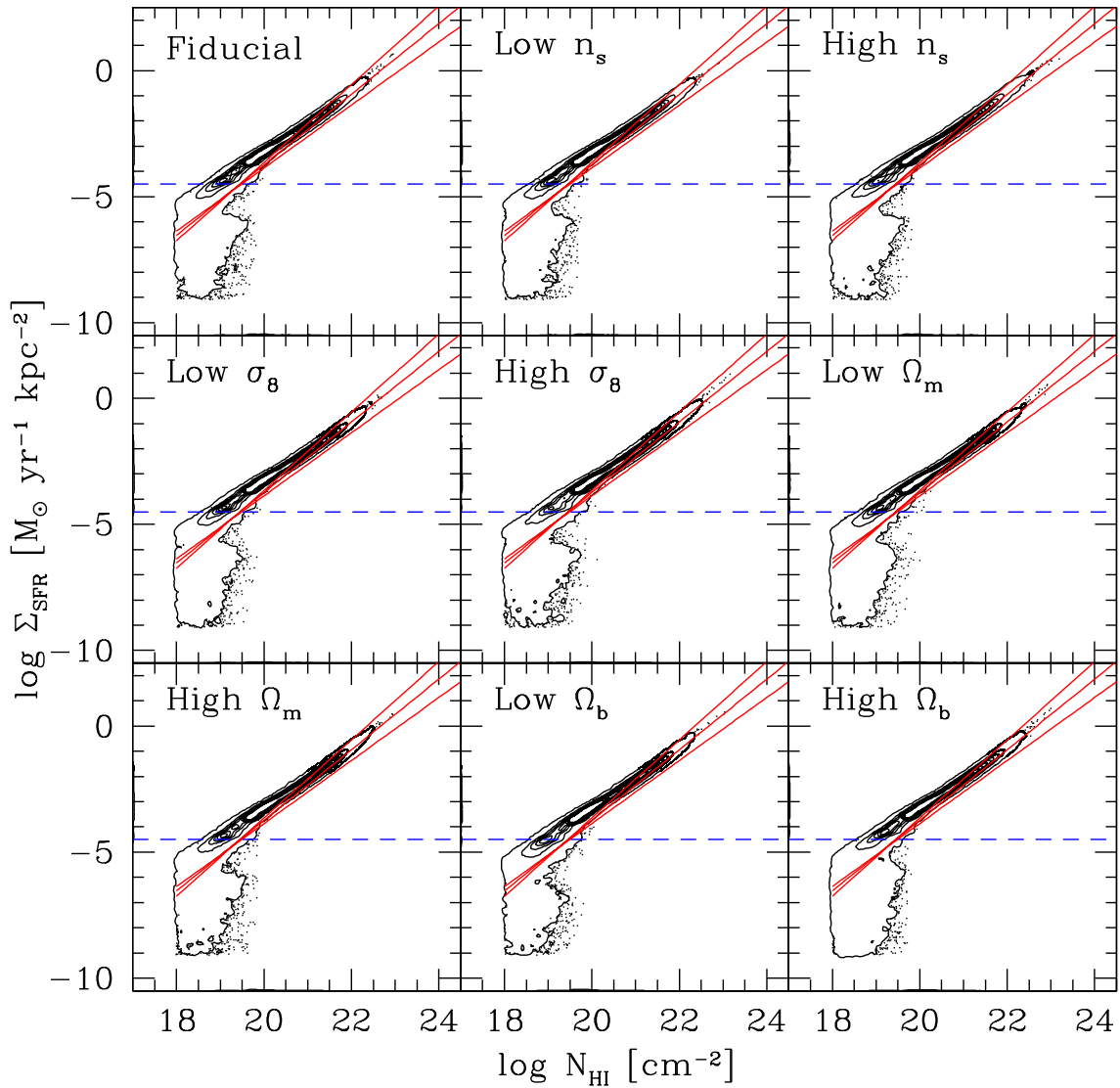}}
\caption{
The projected SFR as a function of \HI\ column density at $z = 3$ for the N216L10 series with different cosmological parameters.
The red line shows the prediction of the Kennicutt-Schmidt law.
The cosmological parameters hardly change the shape of this relation.
The blue dashed line is the current observational limit of the low SFR at $z = 0$.
}
\label{fig:kenn_w5}
\end{figure*}

The variation of $\Omega_{b}$ simply changes the gas density for a given dark matter halo.  
The increase (decrease) of the gas density enhances (reduces) gas cooling.  
Therefore the effect of $\Omega_{b}$ becomes important at lower redshifts when the effect of gas cooling time is more important.
Our simulations confirm that the cosmic SF history is mostly determined by the structure formation at high redshift and by the gas cooling at low redshift, as discussed by \citet{Hernquist.Springel:03}.

Figure~\ref{fig:MF_w5} shows the galaxy stellar mass functions at $z=3$ for the runs with different cosmological parameters. 
The galaxies in the simulation were identified using a simplified variant of the \textup{SUBFIND} algorithm \citep{Springel.etal:01,Choi.Nagamine:09}.
In this paper, we set the mass limit of the simulation galaxies 32 stellar particle mass (see Table~\ref{table:sim2}).
The shape of mass functions is hardly affected by the variation of the cosmological parameters, while the amplitude of them is slightly changed.
One noticeable feature in Figure~\ref{fig:MF_w5} comparing with Figure~\ref{fig:comp_w5} is that the change of mass function owing to the variation of $n_s$ is relatively small.
At first glance, it might appear to be conflicting with the significant change in $\rhosdot(z)$.
However, the SFR in Figure~\ref{fig:comp_w5} is in a logarithmic scale, therefore the seemingly large relative difference in $\rhosdot(z)$ at high-$z$ only results in a small difference in the normalization of the galaxy stellar mass function.
In addition, the difference in $\rhosdot(z)$ caused by $n_s$ becomes negligible at $z$=3.
Generally the qualitative trend in Figures~\ref{fig:comp_w5} and \ref{fig:MF_w5} is the same, with the green dashed line at the top, the red curve in the middle, and the blue dotted line at the bottom.

The cosmological parameters do affect the cosmic star formation history, but they hardly change the SF efficiency as a function of gas column density.
Figure~\ref{fig:kenn_w5} compares the SFR in our simulations with the empirical Kennicutt-Schmidt law (red solid lines, with upper and lower lines showing the errors in Equation~(\ref{equ:KS})). 
All simulations show similar results and relatively good agreement with the Kennicutt-Schmidt law (except for the low column density end, as we will discuss later). 
Figure~\ref{fig:kenn_w5} represents the local star formation relation, while Figure~\ref{fig:comp_w5} represents the global star formation.
This comparison shows that, although the cosmological parameters influence the amplitude of cosmic SFR, they hardly alter the local SFR distribution.
The local star formation is mostly controlled by the adopted SF model. 
In the next section, we will compare the effect of different SF models.

\section{Comparing different star formation models}
\label{sec:psf}

\begin{figure*}
\centerline{\includegraphics[width=0.9\textwidth,angle=0] {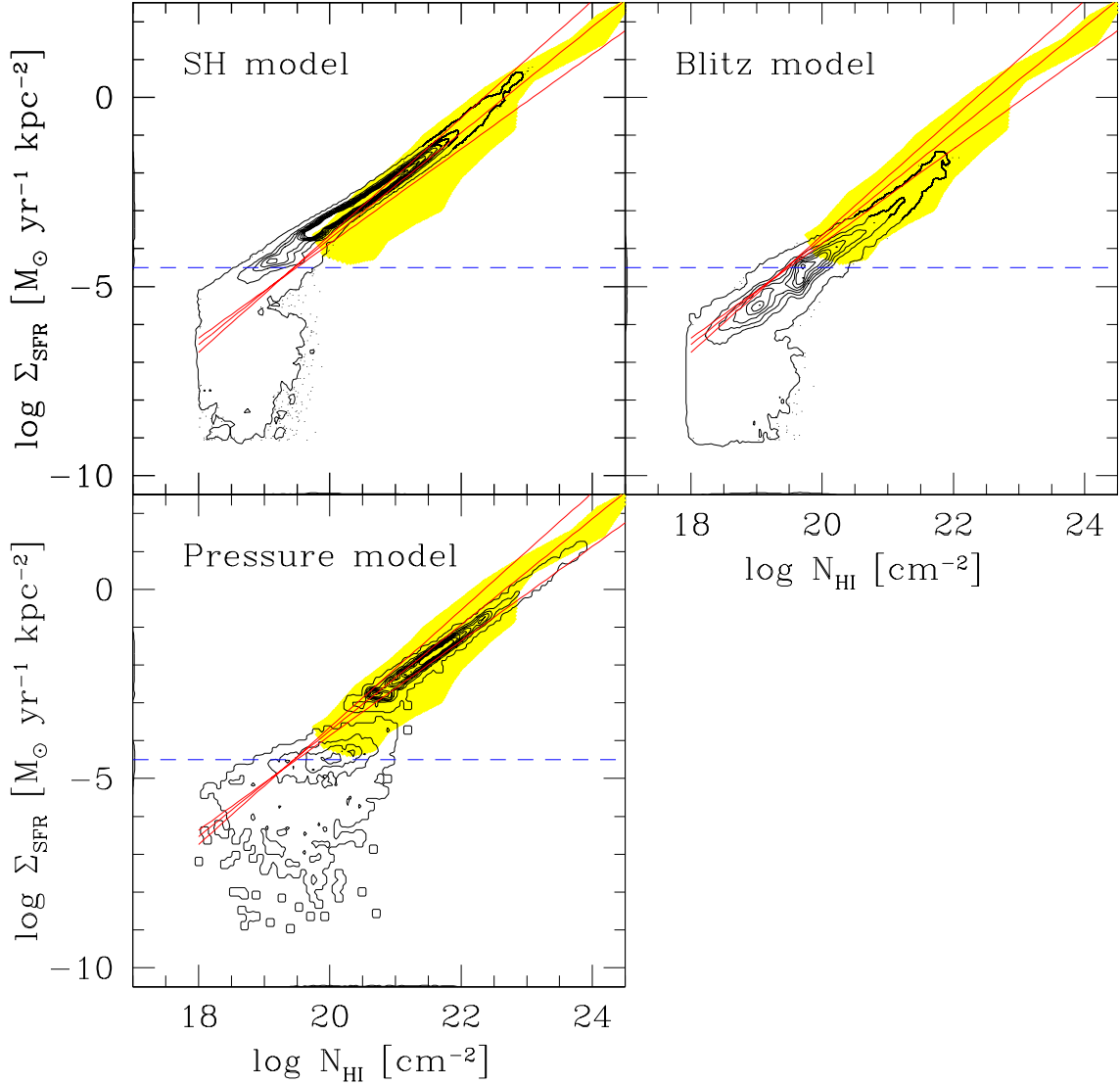}}
\caption{
The projected SFR as a function of \HI\ column density at $z = 3$ for the N216L10 series with different SF models.
The red lines show the empirical Kennicutt-Schmidt law in Equation~(\ref{equ:KS}). 
All three models show reasonably good agreement with the Kennicutt-Schmidt law, however there are notable differences in the three results as we discuss in the text.
The blue dashed line is the current observational limit of the low SFR at $z = 0$.
The yellow shading in each panel indicates the range of observed data for $z = 0$ galaxies taken from the locus of data points in Figure 15 of \citet{Bigiel.etal:08}.
}
\label{fig:kenn_sf}
\end{figure*}

In Section~\ref{sec:sfmodel}, we proposed two new SF models: the Blitz model and the Pressure model.
In this section, we compare the results of these models with those of the original SH model.
Figure~\ref{fig:kenn_sf} shows the projected SFR as a function of H\,{\sc i} column density, which we computed following the prescriptions as described in \citet{Nagamine.etal:04b}.  
The size of the projected pixels is equal to the gravitational softening length in the simulation.  
The three SF models generally agree with the Kennicutt-Schmidt law, but there are some notable differences between the models particularly in the low density and low $\Sigsfr$ regions, which we now discuss.

The projected SFR from simulation is for $z = 3$ result, while the observed data is for $z = 0$.
Most high-$z$ galaxies are currently not resolved well, therefore it is difficult to measure the Kennicutt-Schmidt law for high-$z$ galaxies.
There are some indirect inferences, such as the work by \citet{Wolfe.Chen:06}, that suggest lower star formation efficiencies at high-redshift, but the uncertainties are still very large. 
On the contrary, another work by \citet{Bouche.etal:07} actually suggests the opposite with a four times more efficient star-formation for $z \sim 2$ star-forming galaxies.
Therefore, at this moment, it would be best to compare our simulation results against the most robust local observations of the Kennicutt-Schmidt law.

First, all simulations including those in Figure~\ref{fig:kenn_w5} show a large population of low $\Sigsfr$ with $\log \Sigsfr < -5$ and $\log \NHI \sim 18-19$.
These low $\Sigsfr$ SF regions are below the lower limit of the current observations \citep[e.g.,][]{Bigiel.etal:08}, which is indicated by the dashed horizontal line.
The fraction of star formation occurring in these low $\Sigsfr$ regions are 0.017, 2.9, and 0.0052\% for the SH, Blitz, and Pressure model, respectively, while the number fractions of columns in these low $\Sigsfr$ regions are 18, 72, and 31\% for the SH, Blitz, and Pressure model, respectively.
The total SFR in these low $\Sigsfr$ regions is negligible, because the largest contribution comes from high density and high SFR regions.
However, there is a significant number of columns with these low $\Sigsfr$.
It is possible that the current observations are missing the above fractions of SFR and the number of star-forming columns in the Universe. 
These low $\Sigsfr$ regions might be related to the outskirts of the disk, or the destruction of dwarf galaxies and the formation of tidal tails. 
It would be an interesting future topic of research to further investigate these low $\Sigsfr$ regions in our simulations.  

Second, the SH model overpredicts the $\Sigsfr$ at $\log \NHI \lesssim 20.5$, which is also true for all the runs shown in Figure~\ref{fig:kenn_w5}. 
This feature was also noted by \citet{Nagamine.etal:04b}.
This overprediction of $\Sigsfr$ is absent in the Blitz model and the Pressure model. 
The Blitz model considers the effect of ${\rm H_2}$ on the SFR in the low-pressure regions with low SFRs, and the lower SFRs in the low $\NHI$ regions are expected in this model. 
This appears to be a favorable improvement of the SF model, because \citet{Blitz.Rosolowsky:06} compared the observed $\Sigsfr$ against their SF model and the Kennicutt-Schmidt law for several local galaxies, and found that their model shows better agreement with the observations than the Kennicutt-Schmidt law at the outskirts of disks, where the gas density is lower.
In our cosmological simulations, the Blitz model gives overall lower $\Sigsfr$ than the observed Kennicutt-Schmidt law (but still within the observed range shown by the yellow shading) as shown in the top right panel of Figure~\ref{fig:kenn_sf}.
There might still be some room for improvement of this model in our simulations. 

Third, the Pressure model shows more favorable features compared to both the SH and the Blitz model (see Figure~\ref{fig:kenn_sf}).  
As described in Section~\ref{sec:sfmodel}, the Pressure model adopts a higher SF threshold density than the SH model, and this has two effects. 
One is the reduction of $\Sigsfr$ in low $\NHI$ regions compared to the SH model, and the other is the shift of the SF cut-off column density towards higher value at around $\log \NHI \sim 19.5$. 
Since the current observations only probe down to $\log \Sigsfr \simeq -4.5$, the turn-down may appear to occur at $\log \Sigsfr \simeq 20-21$ \citep[e.g.,][]{Kennicutt:98a, Bigiel.etal:08}.
The changes seen in the Pressure model are favorable improvements over the SH and Blitz models, as they bring the simulation results closer to the observations.
In addition, the $\Sigsfr$ for a given $\NHI$ is lower than the Kennicutt-Schmidt law in the Blitz model, which makes the Blitz model somewhat unfavorable.  
This may be solved by increasing the time-scale parameter $\epsilon$ in the model, however, here we choose not to change the parameter values suggested by the original authors, because they determined those values by observations. 
For these reasons, we favor the Pressure model over the SH and Blitz models. 

Lastly, the slope of $\Sigsfr$ in the Pressure model is slightly shallower than the SH and the Blitz model at high-$\NHI$ with $n\simeq 1.3$
(see Equation~(\ref{equ:KS})).
This is because the Pressure model adopts a different EOS for the cold gas with a smaller value of $\gamma_{eff} = 4/3$ than the usual $\gamma = 5/3$. 
As we mentioned in Section~\ref{sec:sfmodel}, the value of $\gamma_{eff} = 4/3$ has other benefits that it can prevent the artificial fragmentation of gas.
Although the slope is slightly shallow, it is still in the acceptable range of current observational uncertainties.

\section{Cosmic star formation history and galaxy evolution in the new star formation model}
\label{sec:new}

\begin{figure}
\centerline{\includegraphics[width=1.0\columnwidth,angle=0] {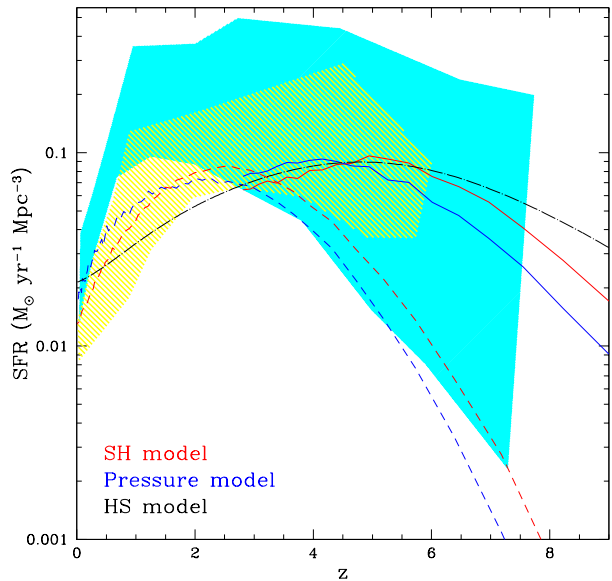}}
\caption{
The cosmic star formation history from our simulations with different SF models.
The solid lines are from the N216L10 series, and the dashed lines are from the N400L100 series.
The N216L10 series represent the high-$z$ SFR better, and the N400L100 series represent the low-$z$ SFR better.
We compare our results with the previous theoretical model of \citet[][the HS model; blue long-dashed line]{Hernquist.Springel:03}.
The cyan shading is the observed range of SF history from \citet{Kistler.etal:09}.
The yellow shading is the locus of the observed data compiled by \citet{Nagamine.etal:06}.  Both compilations of data considered the dust extinction correction.
This figure shows that the peak of the SFR density shifts to a lower redshift in the Pressure model compared to the SH model.
}
\label{fig:comp_sf}
\end{figure}

In this Section, we compare the results of the SH model and the Pressure model 
on the cosmic SF history and galaxy evolution.
Figure~\ref{fig:comp_sf} shows the cosmic SF history for both models.
Here, in order to cover the entire history, we include the results from a larger volume but lower resolution simulation, which are initially made up of $2\times 400^3$ gas and dark matter particles in a $100 h^{-1}$\,Mpc comoving box down to $z=0$ (the N400L100 series). 
The combination of N216L10 and N400L100 series allows us to cover both high and low redshifts, and alleviate any resolution effects. 

At early times, the Pressure model shows lower SFR than the SH model. 
Since the two simulations have the same cosmology and the same initial conditions, the dynamical evolution of gas is identical until the nonlinear growth and star formation is initiated.
As we discussed in Sections~\ref{sec:sfmodel} and \ref{sec:psf}, the star formation is suppressed in the low-density regions in the Pressure model compared to the SH model, and it also has a higher density threshold for star formation, resulting in the suppression of star formation at early times.

\subsection{Peak Redshift}
More interestingly, the suppression of early star formation shifts the peak of $\rhosdot(z)$ to a lower redshift.
In the Pressure model, the peak moves to a lower redshift by $\Delta z = 1-2$ compared to the one in the SH model. 

The total amount of stellar masses at the end of the simulations are similar in the two models owing to the same cosmology and the same structures. 
The SH model forms more stars at high-$z$, but the SF slows down earlier because the total amount of cold gas is limited. 
In contrast, the Pressure model suppresses the early star formation and leaves more cold gas available at lower redshifts, therefore the SFR slows down later than in the SH model. 
The shift of the peak in $\rhosdot(z)$ results from the combined effect of suppression of the early star formation and the limited amount of cold gas at late times. 
Based on Figure~\ref{fig:comp_sf}, we expect that the true peak is located in-between the peaks of N216L10 and N400L100 runs, i.e, at around $z=2-4$. 

The location of the peak redshift of the cosmic SFR has shown considerable discrepancy between observations and theories.
\citet{Springel.Hernquist:03_SFR} claimed that the peak of $\rhosdot(z)$ lies earlier than $z=5$, while observations suggest that the peak is at $2 \lesssim z \lesssim 4$ \citep[e.g.,][]{Hopkins.Beacom:06}.
\citet{Nagamine.etal:04} compared two different types of hydrodynamic simulations, SPH and Eulerian TVD codes, and found a good agreement in $\rhosdot(z)$ between the two simulations with the peak being at $z\ge 4$. 
\citet{Nagamine.etal:06} also compared the predicted stellar mass densities $\rhostar$ with observations, and found that theory predicts higher $\rhostar$ than the observational estimates.  
They both concluded that the early SF seems to be a generic feature of the concordance $\Lambda$CDM model, and suggested that the current observations could be missing nearly half of the $\rhostar$ in the Universe at high-$z$.

In Figure~\ref{fig:comp_sf}, we also compare our simulation results with the theoretical fitting model of \citet[][the HS model]{Hernquist.Springel:03}, which is based on a series of cosmological SPH simulations combined with semi-analytic arguments.
Since our simulations include metal cooling, the star formation is enhanced by $20 - 50$\% compared to the simulations without metal cooling \citep{Choi.Nagamine:09}.
To include this enhancement, we multiply a factor of 1.3 to the original HS model formula in Figure~\ref{fig:comp_sf}.
As expected, the HS fitting agrees well with the SH model.

We also show two compilations of observational estimates of $\rhosdot(z)$
in Figure~\ref{fig:comp_sf}.
The cyan shading is from \citet{Kistler.etal:09}, which combines the data inferred from the \emph{Swift} gamma-ray bursts (GRBs), the data compiled by \citet{Hopkins.Beacom:06}, and the high-$z$ UV data from \citet{Bouwens.etal:07,Bouwens.etal:08}.
The high-$z$ SFRs inferred from GRBs tend to be higher than the previous estimates, which makes the cyan shading quite wide.
The yellow shading shows the observational estimates complied in \citet{Nagamine.etal:06}, who used different dust corrections from \citet{Hopkins.Beacom:06}.
Owing to different dust corrections, the yellow shading is lower at low-$z$, and shows better agreement with our simulations.
The current observations of $\rhosdot(z)$ still show large uncertainties particularly at high-$z$, and seem to suggest that the peak of $\rhosdot(z)$ lies at $z \leq 4$, which is at a lower redshift than the SH model prediction.

The peak of $\rhosdot(z)$ will be better determined in the near future using the data from the next generation of telescopes, such as the 30 meter telescopes and the {\it James Webb Space Telescope} ({\it JWST}).  
These telescopes will measure the faint-end slope of the luminosity function more accurately, and the estimates of the UV luminosity density from star-forming galaxies will become more accurate.   
Our current results suggest that the adopted SF model was partially responsible for the early peak of $\rhosdot(z)$ in the simulations of \citet{Springel.Hernquist:03_SFR} and the HS model. 
We find that the improved SF model can mitigate the conflict that was found by \citet{Nagamine.etal:04} as we describe below.

\begin{figure*}
\subfigure[All stars]{\includegraphics[width=0.48\textwidth,angle=0] {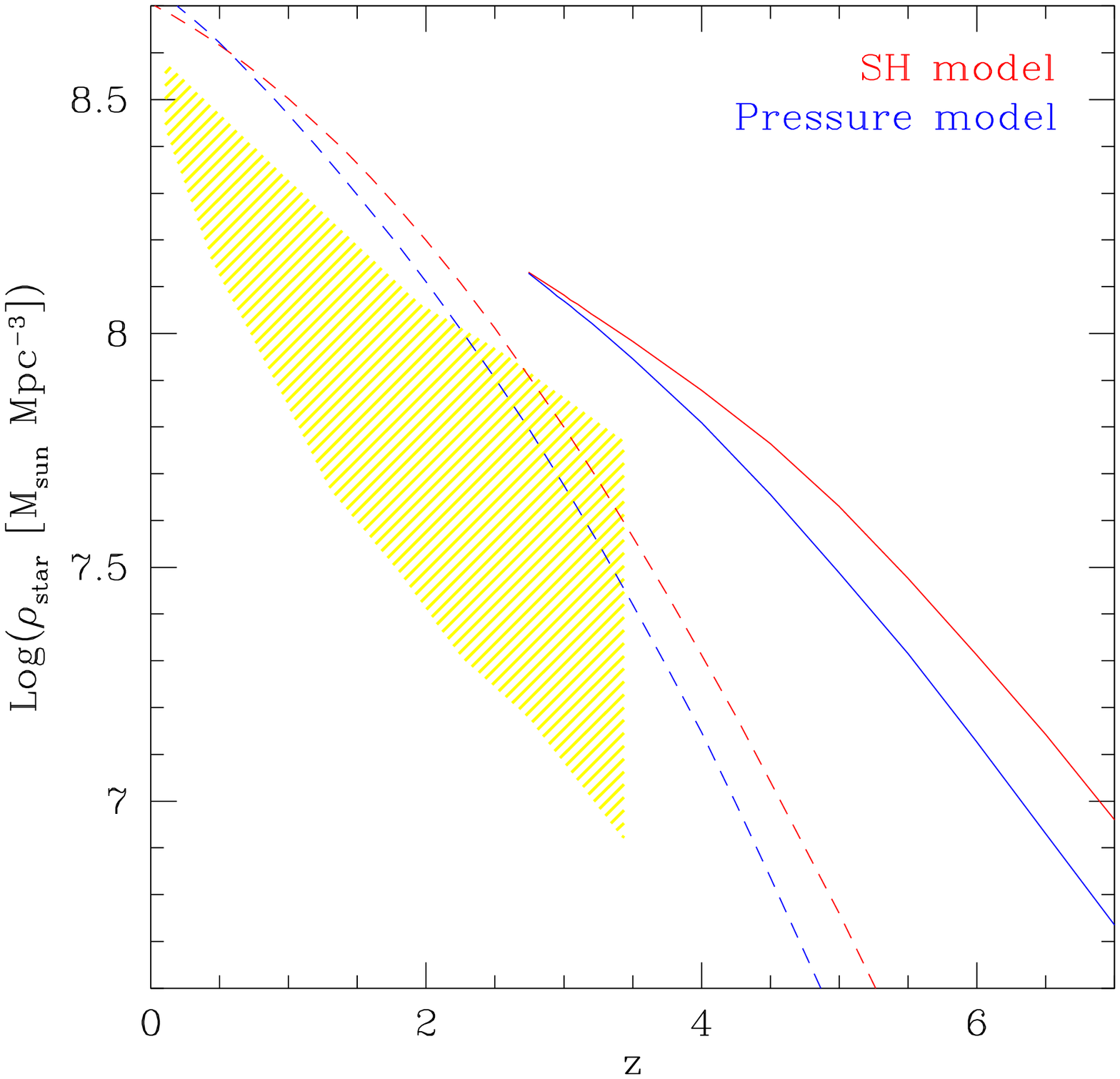}}
\subfigure[Stars in the galaxies with  $\Mstar > 10^8 \Msun$]{\includegraphics[width=0.48\textwidth,angle=0] {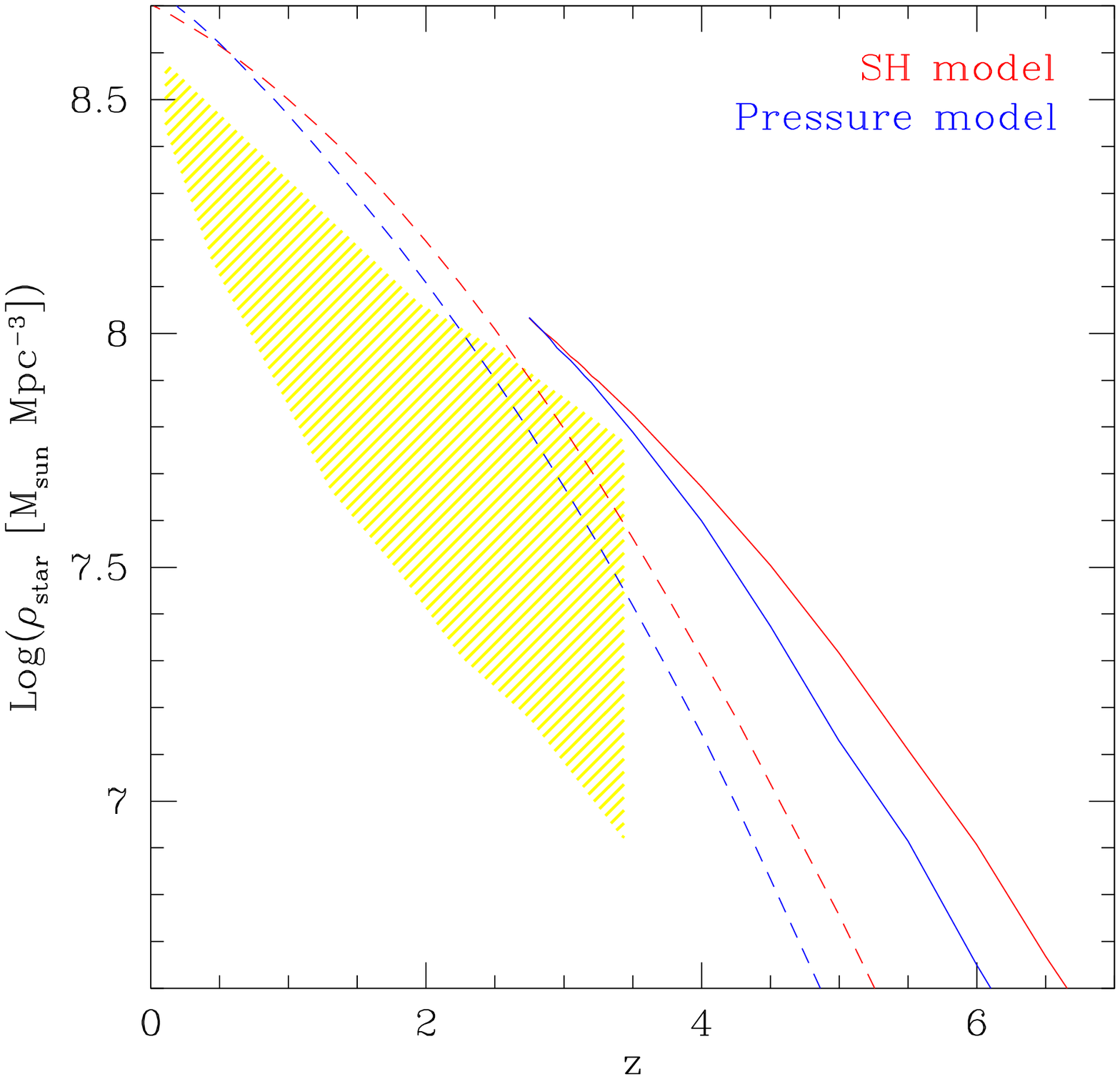}}
\caption{
Evolution of the global stellar mass density $\rhostar$ in the simulations as a function of redshift.
{\it Left panel:} The stellar mass includes all the star particles in the simulations.
{\it Right panel:} The stellar mass is estimated by integrating the stellar mass function only over the range of $\Mstar > 10^{8}\Msun$.
The solid lines are for the fiducial run of N216L10 series, and the dashed lines are for the N400L100 series. The results of the two models are shown (the SH model in red and the Pressure model in blue) for each simulation. 
The yellow shading represents the range of observational estimates shown in \citet{Marchesini.etal:08}.
The SH model predicts higher $\rhostar$ owing to the early star formation. 
The estimate with $> 10^{8} \Msun$ mass-cut (panel $b$) misses the stellar mass density at high redshift.
}
\label{fig:RhoStar}
\end{figure*}

\subsection{Stellar Mass Density}
The delayed star formation in the Pressure model gives rise to changes in the other observables as well.
Figure~\ref{fig:RhoStar} shows the evolution of the global stellar mass density $\rhostar$ as a function of redshift. 
Several observations have estimated the evolution of $\rhostar(z)$ \citep[e.g.,][]{Dickinson.etal:03,Rudnick.etal:06,Perez_Gonzalez.etal:08,Marchesini.etal:08}.
In this figure, we show the range of observational data compiled by \citet{Marchesini.etal:08} with a yellow shading, obtained by integrating the observed stellar mass function.
Note that they fit the observed data at $\Mstar > 10^{10}\Msun$ with a Schechter function by a maximum likelihood method, and the integration down to  $\Mstar = 10^{8}\Msun$ is based on the extrapolated Schechter mass function fit. 
Compared to these observational estimates, our simulations predict somewhat higher $\rhostar$.
Since the $\rhostar$ in the Pressure model tends to be lower than in the SH model, it shows better agreement with the current observational estimates. 

It is still possible that the current observations miss a considerable number of low-mass galaxies at high-$z$, as discussed by \citet{Nagamine.etal:04}. 
To quantify the missed fraction of $\rhostar$, we plot $\rhostar(z)$ with and without the mass-cut of $\Mstar > 10^8 \Msun$ in Figure~\ref{fig:RhoStar}b.
Comparison of the two estimates shows that the measurement with the mass-cut underestimates $\rhostar$ particularly at high-$z$, e.g., by $\sim$65\% at $z=6$. 
This difference is evident only in the N216L10 series, whose galaxy resolution reaches $ < 10^{8} \Msun$.
In the N400L100 series, the galaxy resolution is above $10^8\Msun$, therefore the discrepancy between the two estimates does not show up in Figure~\ref{fig:RhoStar}b. 
Since most of current $\rhostar$ is computed with galaxies with $\Mstar>10^8\Msun$, they could be missing as much as 65\% at $z=6$ and 30\% at $z=3$.
If we change the mass limit to $\Mstar=10^{10}\Msun$, which is the flux limit of current high-$z$ galaxy surveys, instead of $10^8\Msun$, then the missed stellar mass fraction would be more than 50\% at $z=3$. 

Finally, we point out that there is an interesting contradiction between Figure~\ref{fig:comp_sf} and Figure~\ref{fig:RhoStar}.
The observed $\rhosdot(z)$ tends to be higher than our simulations except at very high-$z$, while the observed $\rhostar(z)$ tend to be lower than the simulations.
This contradiction, which was also discussed by \citet{Nagamine.etal:06}, may result from the missed low-mass galaxies at high-$z$ or the redshift evolution of the stellar initial mass function \citep{Dave:08, vanDokkum:08, Wilkins.etal:08}. 
Future observations by 30\,m telescopes and JWST may provide an answer to this conundrum.

\subsection{Baryonic Mass Functions}
\begin{figure*}
\centerline{\includegraphics[width=0.9\textwidth,angle=0] {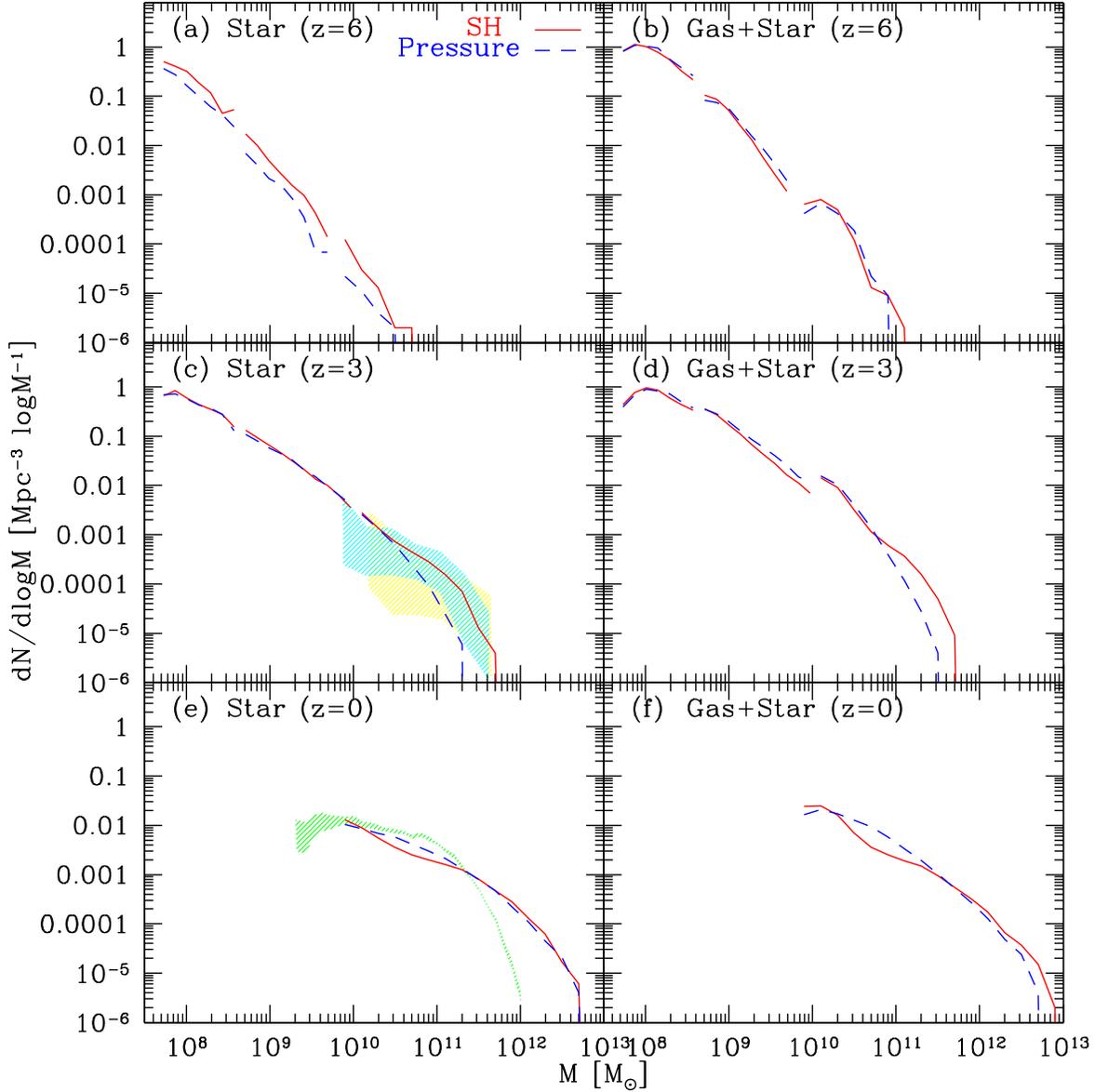}}
\caption{
Baryonic mass functions for the two star formation models: the SH model (solid lines) and the Pressure model (dashed lines).  
The left column panels show the stellar mass functions, and the right column panels show the baryonic (gas $+$ star) mass functions.
At $z=6$ and $z=3$, in each panel we plot three different sets of simulations with different resolution to cover a wide mass range: the N216L10 runs cover the low-mass end, the N400L34 runs cover the intermediate mass range, and the N400L100 runs cover the most massive end of the mass functions.
At $z=0$, we only show the N400L100 run, because the runs with small volumes are stopped at higher redshifts.
The shaded regions at $z=3$ represent the range of observed stellar mass functions at $3<z<4$ (yellow shading) and $2<z<3$ (cyan shading) from \citet{Marchesini.etal:08}.
The green shading at $z=0$ shows the observed local stellar mass function from \citet{Cole.etal:01}. 
The two SF models show different mass functions at $z=6$, but this discrepancy reduces at $z=3$ and $z=0$. 
}
\label{fig:MF_sf}
\end{figure*}

Different cosmic star formation histories may result in different galaxy evolution.
We compare the baryonic mass functions for two different SF models in Figure~\ref{fig:MF_sf}.
At $z=6$, the stellar mass functions (left column) for the two models show noticeable differences: the galaxies in the Pressure model tend to have lower {\it stellar} masses (or the number of galaxies for a given $\Mstar$ is lower). 

However, it does not mean that the total baryonic mass of galaxies is lower in the Pressure model.
The galaxy baryonic mass functions (right column) show that the galaxies in the Pressure model run are sometimes even more massive than those in the SH model run.
This is because the star formation is suppressed in the low density regions in the Pressure model run with a higher SF threshold density, which results in lower SFRs at early times compared to the SH model run. 
The subsequent supernova feedback is also weaker in the Pressure model, leading to more efficient gas accretion. 
Consequently, the galaxies in the Pressure model run can have higher baryonic masses than those in the SH model at early times.

The difference seen at $z$=6 in the stellar mass functions between the two SF models decreases at low redshift.
At $z=3$, the two SF models show similar stellar mass functions except at the very massive end.
The SH model has a larger number of massive galaxies than the Pressure model, and agrees better with the observational data (cyan and yellow shadings).
The larger number of massive galaxies in the SH model can be ascribed to the merger of smaller galaxies that formed earlier more efficiently than in the Pressure model. 
The Pressure model cuts across the observed range, and is at the higher and lower edge of the observed range at $\Mstar\sim 10^{10}\Msun$ and $10^{11.2}\Msun$, respectively.

At $z=0$, the two SF models show similar mass functions.
Both models agree well with the observation at $\Mstar < 10^{11} \Msun$, but they both overpredict the observed data significantly at the massive end. 
This discrepancy may be due to the lack of AGN feedback in our simulations, which is considered to be the major mechanism to quench star formation in massive galaxies \citep[e.g.,][]{Croton.etal:2006}.
We plan to include the evolution of supermassive black holes and their feedback effects in our future simulations.

\subsection{Gas Fraction}
\begin{figure*}
%%\centerline{\includegraphics[width=0.9\textwidth,angle=0] {GFrac_evol_psf.eps}}
\centerline{
 \subfigure[z=6 (N216L10)]{\includegraphics[width=0.9\columnwidth]{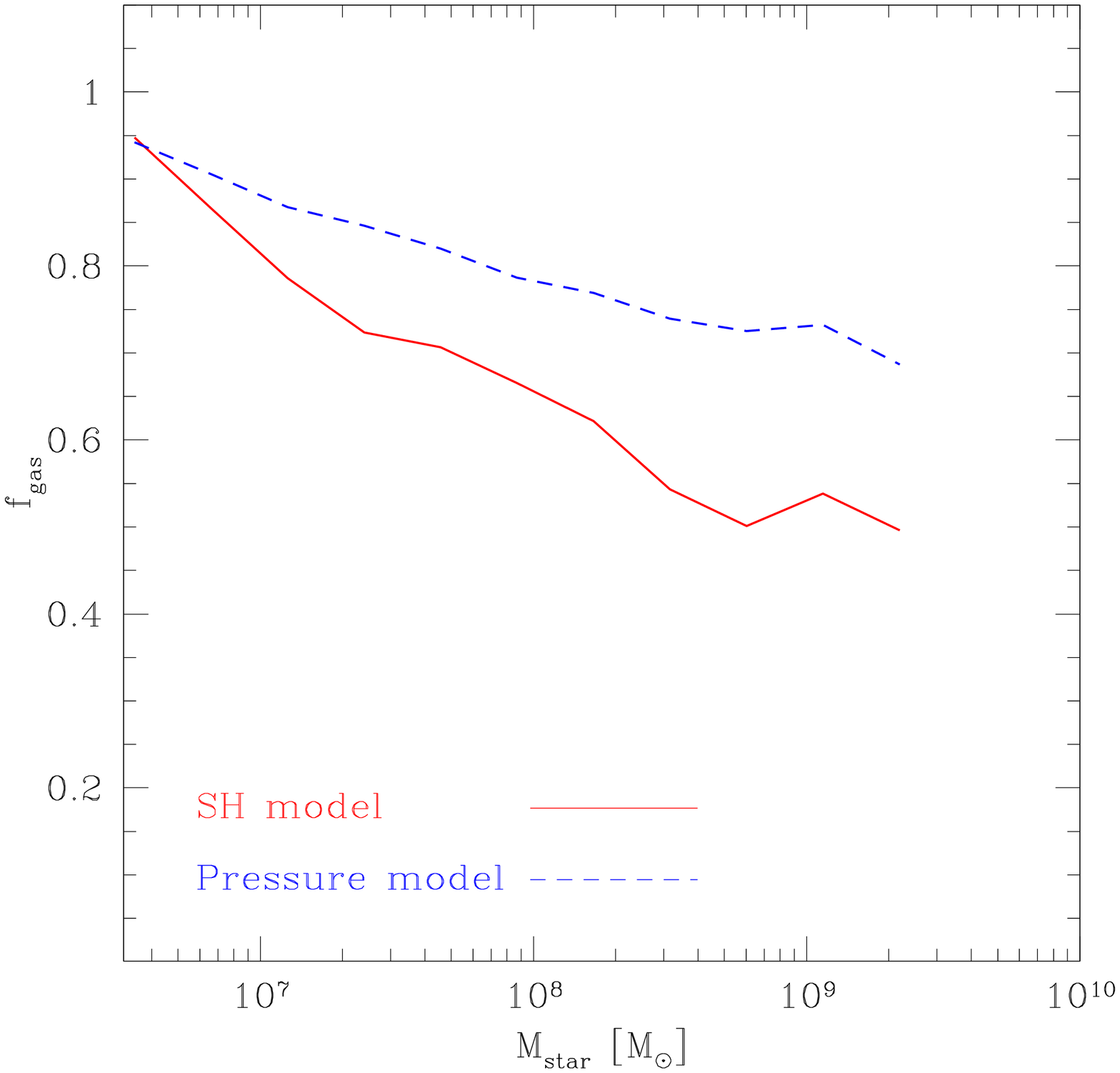}}
 \subfigure[z=3 (N216L10)]{\includegraphics[width=0.9\columnwidth]{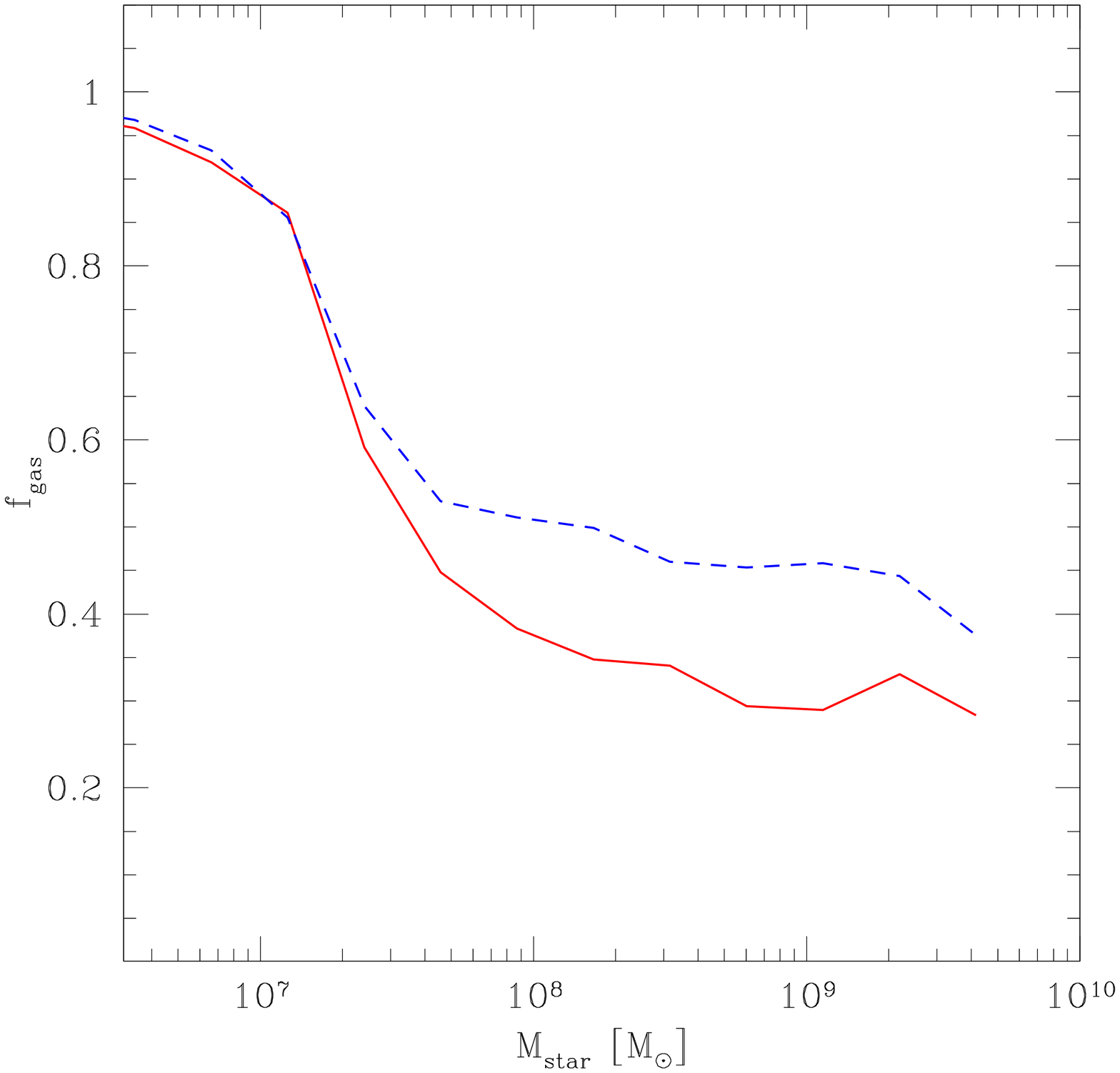}}
}
\centerline{
 \subfigure[z=1 (N400L34)]{\includegraphics[width=0.9\columnwidth]{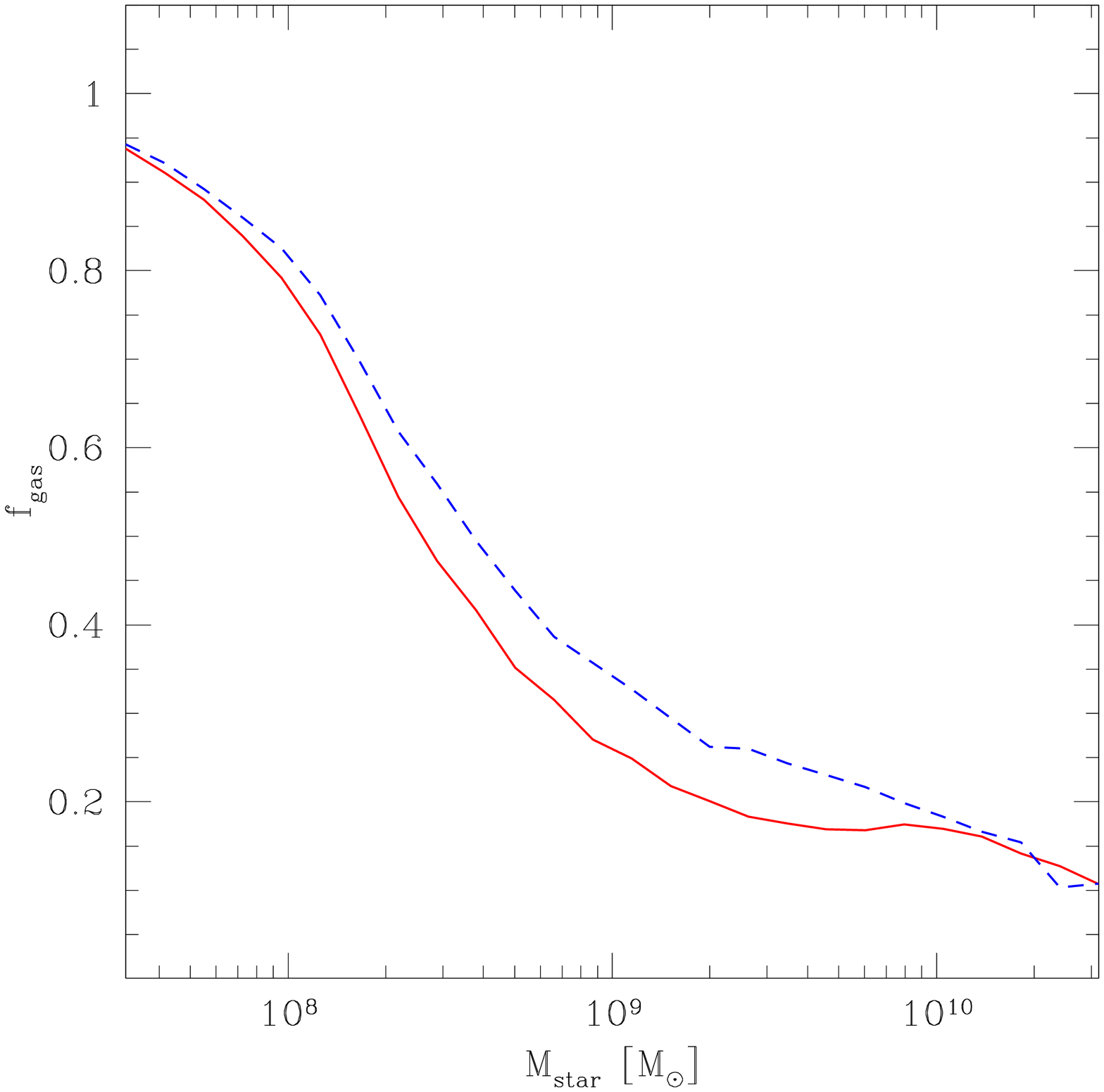}}
 \subfigure[z=0 (N400L100)]{\includegraphics[width=0.9\columnwidth]{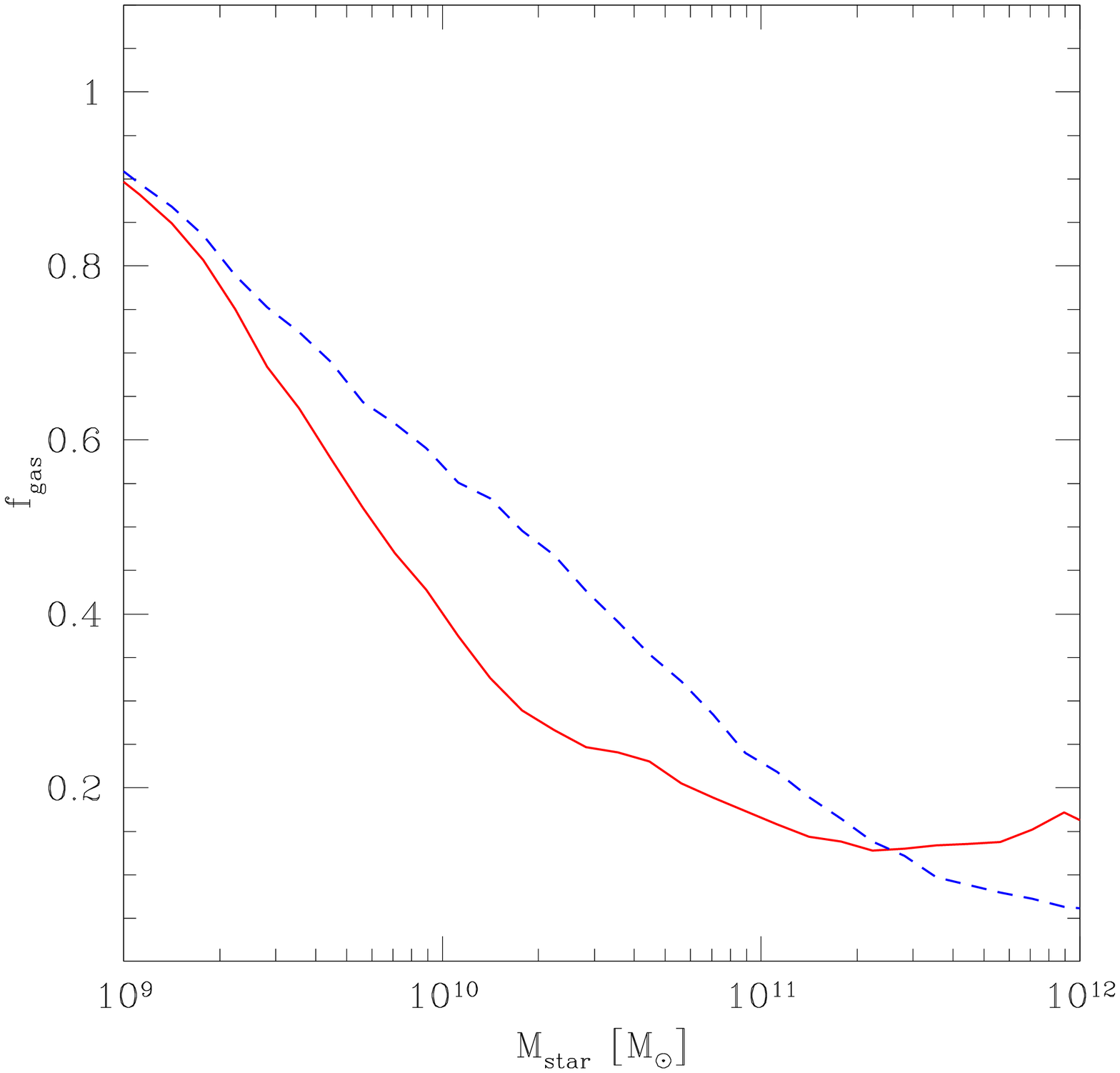}}
}
\caption{
The mean gas fraction of galaxies as a function of galaxy stellar mass.
The galaxies in the Pressure model run are more gas rich than those in the SH model run at all redshifts, except for the most massive galaxies at $z=0$.
}
\label{fig:GF}
\end{figure*}

The suppression of early star formation increases the gas fraction in galaxies.
Figure~\ref{fig:GF} shows the mean gas fraction of galaxies as a function of galaxy stellar mass for the two SF models.
The mean is defined as the ratio of the total gas mass to the total baryon mass (gas$+$stars) for all the galaxies in each mass bin, i.e., $\sum_{i}^{} M_{\rm gas, i} / \sum_{i}^{} M_{\rm baryon, i}$.
The galaxies in the Pressure model run are more gas rich than those in the SH model run at all redshifts, except for massive galaxies at $z=0$.
The balance between galactic outflow and gas accretion determines the gas fraction of the massive galaxies at $z=0$.
We will discuss the effects of different galactic outflow models in 
a separate publication.
Overall, the gas fraction decreases with decreasing redshift. 

One might expect that the increased gas fraction in the Pressure model may enhance the star formation associated with the mergers of gas-rich spiral galaxies.
Recent simulations and models show that the SFR originating from starburst during gas-rich galaxy mergers is  $ \sim 10$\% of the total spheoid mass \citep{Cox.etal:08,Hopkins.etal:09,Hopkins.Hernquist:10}.
Therefore, we expect that the increase of gas fractions in the Pressure model would not change the total amount of star formation from mergers in our simulations very much. 

\section{Summary}
\label{sec:summary}

We have used the cosmological N-body/Hydrodynamics simulation to study the effects of different star formation models on the cosmic star formation history and galaxy evolution.  
Our main results are as follows:

\begin{enumerate}

\item The change of cosmological parameters alters the amplitude of the cosmic SFR density, but does not change the shape of the cosmic SF history very much.  
The cosmological parameters related to the primordial power spectrum ($n_s$ and $\sigma_8$) affect early star formation history significantly, while those related to the matter contents ($\Omega_m$, $\Omega_{\Lambda}$, and $\Omega_b$) effect the SF history more at low redshift with a smaller degree. 
We find that the cosmological parameters hardly change the local SFR for a given \HI\ column density. 

\vspace{0.2cm}
\item  We developed two new star formation models, which consider the effects of ${\rm H_2}$.
Both the Blitz model and the Pressure model reduce the overprediction of projected SFR at low $\NHI$.
However, the Pressure model gives a more realistic SF threshold density and $\Sigsfr$ for a given $\NHI$ than the Blitz model, in better agreement with the empirical Kennicutt-Schmidt law. 
Therefore, we treat the Pressure model to be our new fiducial SF model.  We will continue to refine our SF model in the future. 

\vspace{0.2cm}
\item The Pressure model predicts a large number of columns with $\log \Sigma_{\rm SFR}<-4.5$, which are below the threshold of the current observations. 
If our result is correct, then the current surveys of nearby spiral galaxies \citep[e.g., SINGS][]{Kennicutt.SINGS:03} could be missing $\sim$0.005\% of the total star formation and $\sim$30\% of columns with these low SF surface densities. 

\vspace{0.2cm}
\item The Pressure model reduces the SFR in low-density regions, which causes the suppression of early star formation.
Owing to this suppression, the peak of the cosmic SF history is shifted to a lower redshift, making our results to be more consistent with the recent observational estimates.  
The shift of the peak also decreases the global stellar mass density. 
We find that the Pressure model still predicts higher stellar mass density than the current observational estimates at high redshift.  
If our results are correct, then the current observations could be missing as much as 65\% and 35\% of total stellar masses at $z$=6 and $z$=3, respectively, due to the flux limit of the surveys. 
In particular, our simulations contain a large population of low-mass galaxies with $\Mstar \lesssim 10^8 M_{\sun}$ at $z\gtrsim 3$ that are undetected by the current surveys.  

\vspace{0.2cm}
\item We point out an interesting inconsistency between the observational estimates of the cosmic SFR density and the global stellar mass density (Figure~\ref{fig:comp_sf} vs. Figure~\ref{fig:RhoStar}).  
While the simulation results of the SFR density is on the lower side of the observed range, the predicted stellar mass density is larger than the observed range.  
This suggests that there is an inconsistency between the observational estimates of the two quantities.  
The only uncertainty in this argument is the stellar IMF, which changes the amount of recycled gas and stellar luminosity output per unit mass of collapsing gas cloud. 
Some researchers have invoked a top-heavy IMF to solve this problem (see the discussion in Section~\ref{sec:new}).

\vspace{0.2cm}
\item Owing to the suppression of the early star formation, the galaxies in the Pressure model run tend to have lower stellar masses, but similar total baryonic masses to those in the SH model.
Therefore, the galaxies in the Pressure model run tend to be more gas rich than those in the SH model.
\end{enumerate}

In this paper, we mainly focused on the results from the N216L10 and N400L100 series simulations. 
In the future, we will carry out a larger number of simulation series with different resolution and different box sizes to handle the resolution effects better.
However, we expect that our findings will not change very much, because the qualitative changes due to cosmological parameters and the SF models that we discussed in this paper should be robust against resolution changes.  

We have developed new SF models, however, our treatment still does not explicitly incorporate the formation of $\rm H_2$ owing to the resolution limitation.
Recently, a few groups attempted to carry out a very high resolution simulation with radiative transfer and explicit treatment of $\rm H_2$ formation and destruction \citep[e.g.,][]{Robertson.Kravtsov:08,Gnedin.etal:09}.
So far, they have been able to implement this approach only for a single galaxy, but such a simulation certainly provides important physical insights for an improved star formation model.
\citet{Robertson.Kravtsov:08} confirmed that $\Sigsfr$ shows better correlation with $\rm \Sigma_{{\rm H_2}}$, rather than with $\rm \Sigma_{gas}$.
\citet{Gnedin.etal:09} found that the transition from atomic to fully molecular phase depends on the metallicity.
It implies that the threshold density should be a function of gas metallicity in an improved star formation model.
This causes very low star formation efficiency in the low-mass, low-metallicity galaxies, and one can expect further suppression of star formation in high-$z$ galaxies that are not enriched with metals yet. 
In the future, we will attempt to incorporate the effects of H$_2$ more explicitly and improve our SF models by considering the metallicity effects in cosmological simulations.

%% ==================================================================

\section*{Acknowledgements}
This research was supported in part by the National Aeronautics and Space Administration under Grant/Cooperative Agreement No. NNX08AE57A issued by the Nevada NASA EPSCoR program, by the National Science Foundation (NSF) grant AST-0807491, 
and by the NSF through TeraGrid resources provided by the San Diego Supercomputer Center and the Texas Advanced Computing Center.
We also acknowledge the support by the President's Infrastructure Award at UNLV. 
KN is grateful to the hospitality of the IPMU, University of Tokyo, during the summer of 2009, where part of this work was done.

%\bibliographystyle{mn2e}
%\bibliography{MyRef}

\end{document}